\documentclass[twocolumn,prl,showpacs,amsmath,amssymb]{revtex4}

\usepackage{graphicx}
\usepackage{color}
\usepackage{dcolumn}
\usepackage{bm}
\include{epsf}

\begin{document}

\title{Sub-Kelvin Parametric Feedback Cooling of a Laser-Trapped Nanoparticle}

\author{Jan Gieseler$^1$, Bradley Deutsch$^3$, Romain Quidant$^{1,2}$ and Lukas Novotny$^3$}
\affiliation{$^{1}$ ICFO-Institut de Ciencies Fotoniques, Mediterranean Technology Park, 08860 Castelldefels (Barcelona), Spain}
\affiliation{$^{2}$ ICREA-Instituci{\'o} Catalana de Recerca i Estudis Avan\c{c}ats, 08010 Barcelona, Spain}
\affiliation{$^{3}$ Institute of Optics, University of Rochester, Rochester, NY 14627, USA}

\date{\today}

\begin{abstract}
%
%
%
%
We optically trap a single nanoparticle in high vacuum and cool its three spatial degrees of freedom by means of active parametric feedback.  Using a {\em single} laser beam  for both trapping and cooling we demonstrate a temperature compression ratio of four orders of magnitude.
The absence of a clamping mechanism provides robust decoupling from the heat bath and eliminates the requirement of cryogenic precooling.
%
The small size and mass of the nanoparticle yield high resonance frequencies and high Q-factors along with low recoil heating, which are essential conditions for ground state cooling and for low decoherence.
The trapping and cooling scheme presented here  opens new routes for testing quantum mechanics with mesoscopic objects and for ultrasensitive metrology and sensing.
\end{abstract}

\maketitle

The interaction between light and matter sets ultimate limits on the accuracy of optical measurements. Vladimir B. Braginsky predicted that the finite response time of light in an optical interferometer can lead to mechanical instabilities~\cite{braginsky77} and impose limits on the precision of laser-based gravitational interferometers. Later, it was demonstrated that this ``dynamic back-action mechanism" can be used to reduce the oscillation amplitude of a mechanical system and to effectively cool it below the temperature of the environment~\cite{hoehberger04,cohadon99,arcizet06,gigan06,schliesser06,usami12a} 
and even to its quantum  ground state~\cite{chan11a}. In addition to the fascinating possibility of observing the quantum behavior of a mesoscopic system, many applications have been proposed for such systems ranging from detection of exotic forces~\cite{geraci10a,romeroisart11a,manjavacas10} to the generation of non-classical states of light and matter~\cite{chang10a,romeroisart10a}. \\[-2.2ex]

Most of the mechanical systems studied previously are directly connected to their thermal environment, which imposes limits to thermalization and decoherence. As a consequence, clamped systems require cryogenic precooling. 
A laser-trapped particle in ultrahigh vacuum, by contrast, has no physical contact to the environment~\cite{ashkin76a,ashkin77a}, which makes it a promising system for ground state cooling even at room temperatures~\cite{chang10a,romeroisart10a}. Cooling of micron-sized particles to milli-Kelvin temperatures has recently been achieved by applying an {\em active} optical feedback inspired by atom cooling experiments~\cite{li11a}. A particle is trapped by two counter-propagating beams and cooling is performed with three additional laser beams via radiation pressure. 
However, because light scattering leads to recoil heating there is a limit for the lowest attainable temperature. 
Eliminating recoil heating as the limiting factor for ground state cooling  requires considerably smaller mechanical systems, such as single dielectric nanoparticles~\cite{chang10a,romeroisart10a}. 
Here we demonstrate for the first time optical trapping in high vacuum of a fused silica nanoparticle of radius  $R\sim70\,$nm. Additionally, we employ a novel cooling scheme based on the optical gradient force to cool its motional degrees of freedom from room temperature to $\sim 50\,$mK (compression factor of $\sim10^{4}$).\\[-2.2ex]



\begin{figure}[!b]
\includegraphics[height=20em]{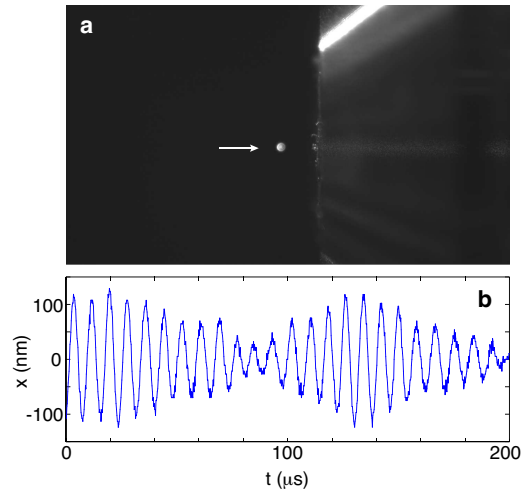}
\vspace{-1em}
\caption{{\bf Trapping of a nanoparticle.}  \textbf{(a)} Photograph of light scattered from a trapped silica nanoparticle (arrow). The object to the right is the outline of the objective  that focuses the trapping laser.  \textbf{(b)} Time trace of the particle's $x$ coordinate (transverse to the optical axis) at $2\,$mBar pressure.
\label{figure1}}
\end{figure}
In our experiments we use a laser beam of wavelength $\lambda=1064\,$nm ($\sim100\,$mW), focused by an NA=0.8 lens mounted in a vacuum chamber. A single nanoparticle is trapped by means of the optical gradient force, which points towards 
the center of the trap for all  translational degrees of the nanoparticle (c.f. Fig.~\ref{figure1}).
For particles much smaller than the wavelength, the polarizability scales as $\alpha\propto R^3$  and the  gradient force dominates over the scattering force. Scattered light from the particle is measured interferometrically with three separate photodetectors that render the particle's motion in the $x$, $y$, and $z$ directions.
This phase-sensitive detection scheme makes use of balanced detection and yields a noise floor of $\sim 1.2 \,{\rm pm/\sqrt{Hz}}$. Fig.~\ref{figure1} shows a photograph of a trapped nanoparticle along with a typical time trace of the particle's $x$ coordinate. Trapping times of more $60$ hours have been achieved at pressures below $10^{-5}\,$mBar indicating that the particle's internal temperature does not affect the center of mass motion 
and that melting of the particle is not a concern.\\[-2.2ex]

To control and stabilize the particle's motion in the optical trap we implemented an active feedback loop.
All three spatial degrees of freedom are controlled with the same laser used for trapping. To cool the center-of-mass motion of the particle we employ a {\em parametric} feedback scheme, similar to parametric amplification of laser fields~\cite{yariv89} and stabilization of nanomechanical oscillators~\cite{villanueva11a}.  After trapping a single nanoparticle at ambient temperature and pressure we evacuate the vacuum chamber in order to reach the desired vacuum level. At ambient pressure the particle's motion is dominated by the viscous force (Stokes force) due to the random impact of gas molecules.   However, as shown in Fig.~\ref{figure1}(b), the inertial force dominates in a vacuum of a few millibars as the particle's  motion becomes ballistic~\cite{li10a}. \\[-2.2ex]

Parametric feedback is activated as soon as we enter the ballistic regime. In a time-domain picture, the feedback loop hinders the particle's motion by increasing the trap stiffness whenever the particle moves away from the trap center and reducing it when the particle falls back toward the trap. In the frequency domain, this corresponds to a modulation at twice the trap frequency with an appropriate phase shift. 
Our parametric feedback is fundamentally different from previous active feedback schemes based on radiation pressure~\cite{poggio07}. Radiation pressure acts only along the direction of beam propagation and therefore requires a separate cooling laser for every oscillation direction~\cite{li11a}. In contrast, the gradient force points towards the center of the trap, thus allowing us to cool all three directions with a single laser beam.\\[-2.2ex]

\begin{figure}[!t]
\includegraphics[width=26em]{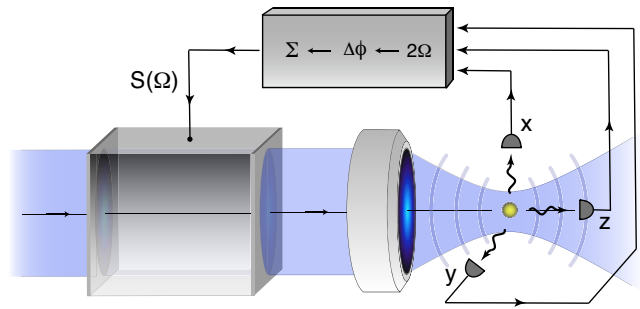}
\caption{{\bf Principle of parametric feedback cooling.} The center-of-mass motion of a laser-trapped nanoparticle in ultrahigh vacuum is measured interferometrically with three detectors, labeled $x$, $y$, and $z$. Each detector signal is frequency doubled and phase shifted. The sum of these signals is used to modulate the intensity of the trapping beam.
\label{figure2}}
\end{figure}

Fig.~\ref{figure2} illustrates our parametric feedback mechanism. To obtain a signal at twice the oscillation frequency we multiply the particle's position $x(t)$ with it's time derivative. The resulting signal $x(t)\,\dot{x}(t)$ is then phase-shifted by a controlled amount in order to  counteract the particle's oscillation.  
Note that depending on the latency of the feedback loop we can achieve damping or amplification of the particle's oscillation. In the absence of active feedback, the particle's oscillation naturally locks to the modulation phase in such a way as to achieve amplification~\cite{yariv89}.
Cooling therefore requires active feedback to adjust the modulation phase constantly.\\[-2.2ex]

\begin{figure}[!b]
\includegraphics[width=27em]{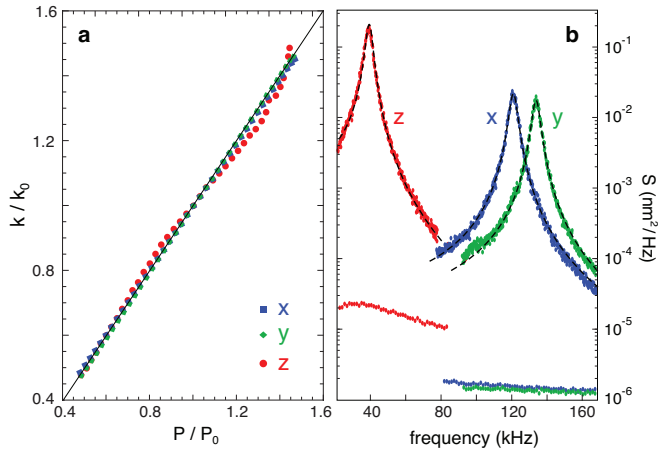}
\vspace{-1.2em}
\caption{{\bf Trap stiffness and spectral densities}. \textbf{(a)}  Normalized trap stiffness in the $x$, $y$, and $z$ directions as a function of normalized laser power. Dots are experimental data and the solid line is a linear fit. \textbf{(b)} Spectral densities of the $x$, $y$, and $z$ motions. The trapped particle has a radius of $R=69\,$nm and the pressure is $P_{\rm gas}=6.3\,$mBar. 
The resonance frequencies are $f_0=37\,$kHz, $120\,$kHz and $134\,$kHz, respectively.
The dashed curves are fits according Eq.~(\ref{eq:SpectralDensity}) and the data on the bottom correspond to the noise floor.
\label{figure3}}
\end{figure}

In our cooling scheme, frequency doubling and phase shifting is done independently for each of the photodetector signals  $x$, $y$ and $z$. Since the three directions are spectrally separated (see Fig.~\ref{figure3}b), there is no cross-coupling between the three signals, that is, modulating one of the signals does not affect the other signals. Therefore, it is possible to sum up all three feedback signals and use the result to drive a single Pockels cell that modulates the power $P$ of the trapping laser. 
 Thus, using a single beam we are able to effectively cool all spatial degrees of freedom.\\[-2.2ex]

For small oscillation amplitudes, the trapping potential is harmonic and the three spatial dimensions are decoupled. Each direction can be characterized by a frequency $\Omega_0$, which is defined by the particle mass $m$ and the trap stiffness $k_{\rm trap}$ as $\Omega_0=\sqrt{k_{\rm trap}/m}$. The equation of motion for the particle's motion in $x$ direction (polarization direction) is 
\begin{equation}
\ddot{x}(t) \,+\, \Gamma_{\!0}\!\; \dot{x}(t) \,+\, \Omega_0^2\;\! x(t)\;=\;  \frac{1}{m}\left[ F_{\rm fluct}(t)  + F_{\rm opt}(t)\right ]   ,
\label{helmhcav5c}
\end{equation}
where $F_{\rm fluct}$ is a random Langevin force that satisfies $\langle F_{\rm fluct}(t)\,  F_{\rm fluct}(t')\rangle = 2 m \Gamma_{\!0\:} k_B T \;\!\delta(t-t')$ according to the fluctuation-dissipation theorem. $F_{\rm opt}(t)=\Delta k_{\rm trap}(t)  \,x(t)$ is a time-varying, non-conservative optical force introduced by parametric feedback. It leads to shifts $\delta\Gamma$ and $\delta\Omega$ in the particle's natural damping rate
$\Gamma_{\!0}$ and oscillation frequency  $\Omega_0$, respectively. Similar equations and considerations hold for the particle's motion in $y$ and $z$ directions. \\[-2.2ex]

We first consider the particle's dynamics with the feedback loop deactivated. For small oscillation amplitudes, the particle experiences a harmonic trapping potential with a trap stiffness $k_{\rm trap}$, which is a linear function of $P$. In the paraxial and dipole approximations (small particle limit, weak focusing) the {\em transverse} trap stiffness is calculated as 
\begin{equation}\label{eqn:ktrap}
k_{\rm trap}\;=\;   4\:\! \pi^3\,\frac{\alpha \:\! P}{c\;\!\varepsilon_o} \frac{{\rm \;NA}^4}{\lambda^4} \, ,
\end{equation}
where NA is the numerical aperture of the focused beam, $\lambda$ is the wavelength, and $\alpha$ is the particle polarizability. A similar expression holds for the {\em longitudinal} trap stiffness. 
For the parameters used in our experiments we find that the particle's oscillation frequency in $x$ direction is $f_0^{(x)}=(k_{\rm trap}/m)^{1/2} /\:\! (2\pi)
=120\,$kHz. For the axial oscillation frequency we find $f_0^{(z)}=37\,$kHz and for the $y$ direction we measure $f_0^{(y)}=134\,$kHz. The different oscillation frequencies in $x$ and $y$ directions originate from the symmetry of the  laser focus~\cite{novotny06b}. The linear dependence of the trap stiffness on laser power has been verified for all three directions and is shown in Fig.~\ref{figure3}(a).
In Fig.~\ref{figure3}(b) we show the spectral densities of the $x$, $y$, and $z$ motions recorded 
at a pressure of $P_{\rm gas}=6.3\,$mBar. \\[-2.2ex]

\begin{figure}[!t]
\includegraphics[width=23em]{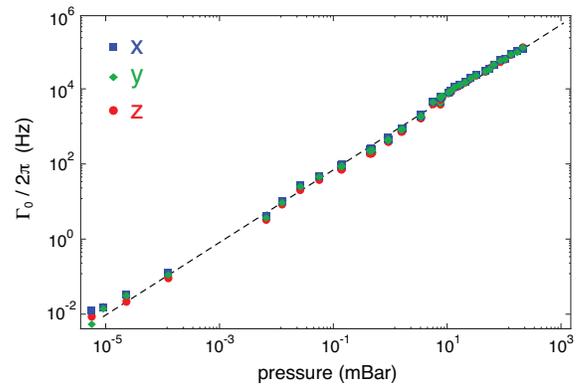}
\caption{{\bf Damping rate as a function of gas pressure.} The damping rate  $\Gamma_0$ decreases linearly with pressure $P_{\rm gas}$. The dashed line is a fit according to Eq.~(\ref{eqn:Gamma0}).
\label{figure4}}
\end{figure}
Once a particle has been trapped, the interaction with the background gas  thermalizes its energy with the environment and, according to the fluctuation-dissipation theorem, damps the particle's motion with the rate $\Gamma_0$ in Eq.~(\ref{helmhcav5c}).  From kinetic theory we find that~\cite{li11a,beresnev90}
\begin{equation}\label{eqn:Gamma0}
\Gamma_0\;=\; \frac{6\pi\eta R}{m}\frac{0.619}{0.619 +{\rm Kn}}(1+c_K)\; ,
\end{equation}
%
where $c_K=0.31 {\rm Kn}/(0.785+1.152{\rm Kn}+{\rm Kn}^2)$, $\eta$ is the viscosity coefficient of air and ${\rm Kn}=\bar{l}/R$ is the Knudsen number. When the mean free path $\bar{l}\propto 1/P_{\rm gas}$ is much larger than the radius of the particle, $\Gamma_0$ becomes proportional to $P_{\rm gas}$. Fig.~\ref{figure4} shows the measured value of $\Gamma_0$ for all three directions as a function of pressure. 
%
%
For a pressure of $P_{\rm gas}=10^{-5}\,$mBar we measure a damping of $\Gamma_0 = 10\,$mHz, which corresponds to an unprecedented quality factor of $Q = 10^{7}$. In ultrahigh vacuum ($P_{\rm gas}=10^{-9}$ mBar), the quality factor will reach values as high as $Q\sim10^{11}$.\\[-2.2ex]

Activation of the  parametric feedback loop gives rise to additional damping $\delta\Gamma$ and a frequency shift $\delta\Omega$. The resulting spectral line shapes are defined by the power spectral density $S_x(\Omega)$, which follows from Eq.~(\ref{helmhcav5c}) as 
\begin{eqnarray}
\label{eq:SpectralDensity}
S_x(\Omega) &=& \int_{-\infty}^{\infty} \!\!\big\langle x(t)\;\!x(t\!-\!t')\big\rangle\, {\rm e}^{- i \Omega t'} \,dt' \\
&=&\frac{\Gamma_{\!0}\,k_B T\,/\,(\pi\:\!m)}{([\Omega_0+\delta\Omega]^2 - \Omega^2)^2 +  \Omega^2 [\Gamma_{\!0}+\delta\Gamma]^2}\, . \nonumber
\end{eqnarray}
Integrating both sides over $\Omega$ yields the mean square displacement
\begin{equation}
 \left\langle x^2\right\rangle \; =\; \big\langle x(0)\;\!x(0)\big\rangle \,= \frac{k_B T}{m (\Omega_0+\delta\Omega)^2}\,\frac{\Gamma_{\!0}}{\Gamma_{\!0}+\delta\Gamma}\; .
\label{helmhcav8}
\end{equation}
According to the equipartition principle, the center-of-mass temperature $T_{\rm cm}$ follows from  $k_{\rm B} \;\!T_{\rm cm}=m\:\!(\Omega_0\!+\!\delta\Omega)^2 \! \left\langle x^2\right\rangle$. Considering that $\delta\Omega\ll \Omega_0$ we obtain
\begin{equation}\label{eqn:Tcm}
T_{\rm cm} \;=\; T \, \frac{\Gamma_{\! 0}}{\Gamma_{\! 0} +\delta\Gamma}\; ,
\end{equation}
where $T$ is the equilibrium temperature in the absence of the parametric feedback ($\delta\Gamma=0$). Thus, the temperature of the oscillator can be raised or lowered, depending on the sign of $\delta\Gamma$ in Eq.~(\ref{eqn:Tcm}).\\[-2.2ex]

\begin{figure}[!b]
\includegraphics[width=26em]{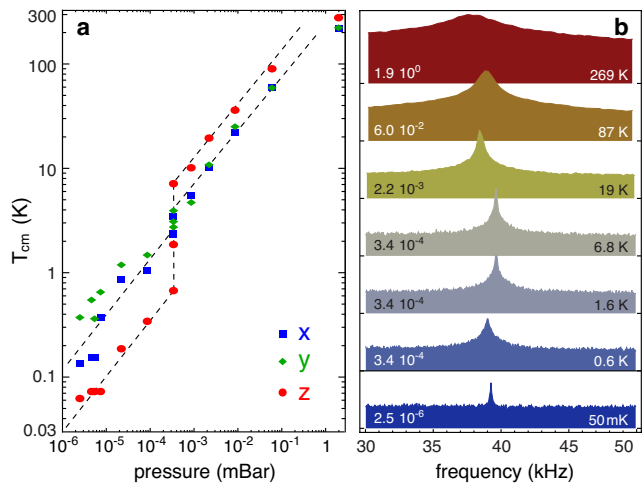}
\caption{{\bf Parametric feedback cooling.} \textbf{(a)} Dependence of the center-of-mass temperature $T_{\rm cm}$ on pressure. The cooling rate (the slope of the dashed lines) is similar for the different directions $x$, $y$ and $z$. The feedback gain has been increased at a pressure of $\sim 0.3\,\mu$Bar causing a kink in the curves. \textbf{(b)} Spectra of the $z$ motion evaluated for different pressures and temperatures $T_{\rm cm}$. The area under the curves is proportional to $T_{\rm cm}$. The numbers in the figure indicate the pressure in mBar.
\label{figure5}}
\end{figure}

The experimental results of parametric feedback cooling are shown in Fig.~\ref{figure5}, which depicts the dependence of the center-of-mass temperature $T_{\rm cm}$ on pressure. 
The cooling action of the feedback loop competes with reheating due to collisions with air molecules, ultimately setting a minimum achievable temperature for each pressure value. 
Since the area under the lineshape  defined in Eq.~(\ref{eq:SpectralDensity}) is proportional to $T_{\rm cm}$,  feedback cooling not only increases the linewidth but also lowers the signal amplitude until it reaches the noise floor. 
Nevertheless,  we are able to reach temperatures of $T_{\rm cm}\sim 50\,$mK while maintaining the particle in the trap.  \\[-2.2ex]

The here introduced trapping and cooling technique represents an important step towards  ground state cooling.
In the quantum limit, a mechanical oscillator exhibits discrete states separated in energy by $\hbar(\Omega_0+\delta\Omega)\sim\hbar\Omega_0$. The mean thermal occupancy is
\begin{equation}
\langle n\rangle = \frac{k_B T_{\rm cm}}{\hbar\Omega_0}  \; .
\label{thermocc}
\end{equation}
In order to resolve the quantum ground state we require $\langle n\rangle < 1$. For a $120\,$kHz oscillator, this condition implies $T_{\rm cm}\sim 6\,\mu$K. According to Eq.~(\ref{eqn:Tcm}), a low pressure implies a low damping rate and thus, extrapolating Fig.~\ref{figure5}a, we find that this temperature will be reached at ultrahigh vacuum ($10^{-11}\,{\rm mBar}$), provided that the particle oscillation can be measured and the feedback remains operational. 
Alternatively, lower occupancy can be reached at higher pressures by an increase of the feedback gain.
Laser power noise introduces fluctuations in the trap stiffness and therefore in the mechanical oscillation frequency. We believe that the resulting random phase error in the feedback loop is the current limiting factor in cooling. 
This phase error can be minimized by  using background suppression and laser stabilization techniques~\cite{seifert06}. The noise  floor in our measurements is currently $1.2 \,{\rm pm /\sqrt{Hz}}$.\\[-2.2ex]

In feedback cooling, the particle's position has to be measured in order  to operate the feedback loop. Measurement uncertainty of $x$, $y$, and $z$ introduced by shot-noise therefore limits the lowest attainable temperature $T_{\rm cm}$. The measurement accuracy is fundamentally limited by the standard quantum limit, which follows from the uncertainty principle $\Delta x\: \Delta p \ge \hbar/2$, where $\Delta p =\Delta n\: \hbar\:\! k$, $\Delta n$ being the uncertainty in photon number and $k=2\;\!\pi\:\!/\:\!\lambda$. For shot noise $\Delta n \propto N^{1/2}$, where $N$ is the mean photon number $N= P\:\! \Delta t\,/\, (\hbar \:\!k \;\!c)$. In terms of the bandwidth $B=1/\Delta t$ we obtain $\Delta x \ge [\hbar\;\! c\;\! \lambda \;\!B \,/\, ( 8 \pi\;\! P)]^{1/2}$. Thus, the measurement uncertainty is determined by the bandwidth $B$ and the signal power $P$ at the detector. 
For a $R\sim70\,$nm nanoparticle and the parameters used in our experiments we find $\Delta x\ge 6.7\,$pm, which corresponds to a center-of-mass temperature of $T_{\rm cm}=7.1\,\mu$K.
Thus, in absence of back action, parametric feedback should allow us to cool a laser-trapped nanoparticle close to its quantum ground state.\\[-2.2ex]


Evidently, the measurement uncertainty $\Delta x$ can be reduced by increasing the signal power at the detector, for example by higher laser power or by using a larger particle size $R$ and hence a larger scattering cross-section  $\sigma_{\rm scatt}=k^4 |\alpha|^2 / (6\pi\varepsilon_o^2)$. However,  strong scattering introduces recoil heating, which destroys the coherent particle motion. 
In analogy to atomic trapping, the transition rate $\Gamma_{\rm recoil}$ between consecutive harmonic oscillator states is calculated as~\cite{chang10a,cirac92}
\begin{equation}
\Gamma_{\rm recoil}\;=\; \frac{2}{5}\,\left[\frac{\hbar k^2\:\!/\:\! 2m}{\Omega_0}\right]\,\left[\frac{I_0\,\sigma_{\rm scatt}}{\hbar\omega} \right] \; ,
\label{recoil}
\end{equation}
where $I_o$ is the laser intensity at the focus. The last term in brackets corresponds to the photon scattering rate. Comparing $\Gamma_{\rm recoil}$ with the frequency of a center-of-mass oscillation $\Omega_0$ we find that in the current configuration there is only one recoil event per $\sim 10$ oscillations. Thus, the trapped nanoparticle can coherently evolve for many oscillation periods.  The number of coherent oscillations in between recoil events $N_{\rm osc}$  scales with the ratio $(\lambda/R)^3$, so small particles and long wavelengths are favorable. \\[-2.2ex]

Our discussion highlights the tradeoff between measurement uncertainty and recoil heating. A nanoparticle of size of $R\sim70\,$nm is a good compromise  between the two limiting factors. Notice that $\Gamma_{\rm recoil}$ and the photon scattering rate differ by a factor of $\sim 10^{-9}$, and hence most of the scattered photons do not alter the center-of-mass state of the particle. The possibility of observing the particle without destroying its quantum coherence is a critical advantage over atomic trapping and cooling experiments. Finally, parametric cooling should work even without continuously tracking  $x(t)$  as long as the frequency and the phase of the center-of-mass oscillation are known.\\[-2.2ex]

In conclusion, we have demonstrated that an optically trapped nanoparticle in high vacuum can be efficiently cooled in all three dimensions by a parametric feedback scheme. The parametric feedback makes use of a {\em single} laser beam and is therefore not limited by alignment inaccuracies of additional cooling lasers. Theoretical considerations show that center-of-mass temperatures close to the quantum ground state are within reach. To fully exploit the quantum coherence of a laser-trapped nanoparticle, parametric feedback cooling can be combined with {\em passive} dynamical back-action cooling~\cite{kippenberg07}, for example by use of optical cavities~\cite{chang10a,verhagen12} or electronic resonators~\cite{teufel11a}. The results shown here also hold promise for ultrasensitive detection and sensing~\cite{geraci10a}. The ultrahigh quality factors and small oscillation amplitudes yield force sensitivities on the order of $10^{-20}\,{\rm N}/\sqrt{Hz}$~\cite{stipe01}, which outperforms most other ultrasensitive force measurement techniques by orders of magnitude, and  can find applications for the detection of single electron or nuclear spins~\cite{rugar04},
Casimir forces and vacuum friction,
phase transitions, and non-Newtonian gravity-like forces~\cite{geraci10a}.\\

This research was funded by the U.S. Department of Energy (grant DE-FG02-01ER15204), by Fundaci{\'o} Privada CELLEX, and ERC-Plasmolight (\# 259196). We thank  Mathieu Juan and Vijay Jain for valuable input and help.\\

Correspondence and requests for materials should be addressed to L.N.~(email:   lukas.novotny@rochester.edu) or R.Q.~(email: romain.quidant@icfo.es)


\end{document}